\documentclass[pdf,epjc3]{svjour3}  
\smartqed  
\RequirePackage{graphicx}
\RequirePackage{mathptmx}      
\RequirePackage[T1]{fontenc}
\RequirePackage{flushend}
\RequirePackage[numbers,sort&compress]{natbib}
\RequirePackage[colorlinks,citecolor=blue,urlcolor=blue,linkcolor=blue]{hyperref}
\usepackage{amsmath,amssymb}
\usepackage{xcolor}
\journalname{Eur. Phys. J. C}
\begin{document}

\title{Mass of compact stars in f(T) gravity}


\author{J. C. N. de Araujo\thanksref{e1,addr1} 
\and 
H. G. M. Fortes\thanksref{e2,addr2} }                     
%
%

\thankstext{e1}{e-mail: jcarlos.dearaujo@inpe.br}
\thankstext{e2}{e-mail: hemily.gomes@gmail.com}

\institute{Divisão de Astrof\'{i}sica, Instituto Nacional de Pesquisas Espaciais, Avenida dos Astronautas 1758, S\~{a}o Jos\'{e} dos Campos, SP 12227-010, Brazil \label{addr1} \and Universidade Federal de Juiz de Fora, Campus Universitário, Rua José Lourenço Kelmer, s/n - São Pedro, Juiz de Fora - MG, 36036-900\label{addr2}}





\date{Received: date / Accepted: date}

\maketitle

\begin{abstract}
{The mass of compact objects in General Relativity (GR), which as is well known, is obtained via the Tolman - Oppenheimer - Volkov (TOV) equations, is a well defined quantity. However, in alternative gravity, this is not in general the case. In the particular case of $f(T)$ gravity, where $T$ is the scalar torsion, some authors consider that this is still an open question, since it is not guaranteed that the same equation used in TOV GR holds. In this paper we consider such an important issue and compare different ways to calculate the mass of compact objects in $f(T)$ gravity. In particular, we argue that one of them, the asymptotic mass, may be the most appropriate way to calculate mass in this theory. We adopt realistic equations of state in all the models presented in this article.}
\end{abstract}

\section{Introduction}
\label{int}

{The mass of compact objects in general relativity (GR) is unambiguously given. However, this is not necessarily the case in some alternative theories of gravity (see, e.g., \cite{olmo}, for a brief discussion). Consequently, it is not possible to probe appropriately such theories against observational data and to compare them with other theories unambiguous.}

{In the so-called Teleparallel Gravity theory \cite{TT}, curvature $R$ is replaced by torsion $T$ and $f(T)$ is a modification of Teleparallel Equivalent of General Relativity (TEGR) \cite{Ferraro} as well as the $f(R)$ is a generalization of the Einstein-Hilbert action. Also, the geometry of spacetime is described in the tetrad-spin connection formulation. For a detailed discussion, see the review paper \cite{Bahamonde}.}

{Different applications have also been investigated in these modified theories of gravity with torsion. The existence of relativistic stars was first investigated in \cite{Bohmer}. Compact stars in a specific $f(T)$ model for an isotropic fluid with a polytropic equation of state were presented in \cite{Ilijic} and latter the authors extended the analysis for boson stars \cite{Ilijic20}. In \cite{paperII,paperIII}, we have investigated the modelling of neutron stars using polytropic equations of state and realistic ones. Recently, a study of neutrons stars also have been considered in \cite{Vilhena} using realistic equations of state. Strong magnetic field effects on neutron stars were addressed in \cite{Ganiou} and pulsars in teleparallel
gravity in \cite{Nashed}. Other alternative models such as $f(R, T, R_{\mu\nu}T^{\mu\nu})$ and $f(T,\mathcal{T})$ have been considered in modelling of compact stars \cite{Sharif,Zubair,Saleem} and quark stars \cite{Pace}.}


{The mass of stars in $f(T)$ is considered by some authors to be still an open problem, see, for example, \cite{Ilijic,Ilijic3}. There is an ambiguity on its definition which brings difficulties in computing it. It is well known that in GR the mass of a compact star can be obtained from $M=4\pi\int_0^R \rho \, r^2 \, dr$ and it is in agreement with the so-called ADM mass \cite{ADM} of the spacetime.}

{The space-time exterior to a non-rotating and spherically symmetric object in GR is given, as is well known, by the Schwarzschild metric. The geometrical mass that appears in this very metric is the mass of the object. From the interior Schwarzschild solution one obtains the Tolman-Oppenheimer-Volkoff (TOV) equations \cite{TOV}. The mass of the object obtained from the interior solution is unambiguously the same that appears in the exterior Schwarzschild metric.}

In $f(T)$ gravity, the vacuum metric is not available in closed form.
Thus, the vacuum space-time is not given by the Schwarzschild metric, but only a perturbative expression in the weak gravity regime was presented in \cite{Ilijic3}. {An approach with non-Schwarzschild vacuum have also been obtained in \cite{Atazadeh}.}

An alternative way of measuring the mass of a star is  related to the total particle number $N$ through the integration of the particle number density $n = dN/dV$ over the interior of the star. This calculation provides the total rest mass $M_0$, although it is not an observable, it can be used to make comparisons between different theories. It is worth noting that the calculation of $M_0$ is unambiguously given, since it only takes into account the total rest mass.  The internal energy and the gravitational potential energy do not enter in the calculation of $M_0$.

The main goal of the present study is to investigate different ways to calculate mass of compact stars in $f(T)$ gravity and identify among them the most appropriated. To do so we consider a particular $f(T)$ gravity, namely, $f(T) = T + \xi T^2$. 
{This particular choice, despite simple, allows us to perform all the required analysis regarding to the mass definition. However, the same analysis can be done for other $f(T)$ models in order to investigate possible additional conclusions and eventual comparisons.}

In Section \ref{be}, the basic equations of $f(T)$ gravity for spherical symmetry are presented, in Section \ref{models}, different ways to calculate mass are presented and compared, identifying the one that may be the most appropriate and, finally, in Section \ref{conclusions} the main conclusions are presented.

\section{The basic equations of f(T) gravity for spherically symmetric spacetime}
\label{be}

{The detailed formulation and the basic equations used in this paper can be found in one of our previous paper \cite{paperI}. Here we present only the main equations.} 

{The equations that come from the action for the $f(T)$  gravity read \cite{paperI}:}
\begin{eqnarray} 
  &\,& 4\pi P = -\frac{f}{4} +\frac{f_T\,e^{-B}}{4r^2} \left( 2-2\,e^B+r^2e^B T-2r\,A' \right) 
\label{026}\\ 
&\,&4\pi\rho = \frac{f}{4} -\frac{f_T\,e^{-B}}{4r^2}\left( 2-2\,e^B+r^2e^B T-2r\,B' \right) 
- \frac{f_{TT}\,T'e^{-B}}{r} \left(1-e^{B/2}\right) 
\label{025}\\ 
  &\,&f_T \Big[ 4\,e^A-4e^Ae^B-e^Ar^2A'^2+ 2\,e^Ar\,B' +e^Ar\,A' \left( 2+r\,B' \right) -2r^2e^A A'' \Big] +\nonumber \\
&\,& +  f_{TT}\Bigl[ -4\,e^A r\, T'(1-e^{B/2} \sin{\gamma})  -2r\,A' e^Ar\,T'\Bigr]=0
\label{027}
\end{eqnarray}
where $P$ and $\rho$ are, respectively, the pressure and the energy density of the star, $f_T=\frac{\partial f}{\partial T}$, $f_{TT}=\frac{\partial f_T}{\partial T}$,  and the prime stands for derivative with respect to the r coordinate. In the above equations, $A$ and $B$ depend only on the $r$ coordinate. {Since we consider spherical stars, the metric can be written as follows}
\begin{equation}
ds^2=e^{A(r)}\, dt^2-e^{B(r)}\, dr^2-r^2\, d\theta^2-r^2 \sin ^2 \theta \, d\phi^2.
\label{metric}
\end{equation}

Inspired in the Starobinsky model in $f(R)$, we consider in this paper the following $f(T)$ gravity model,
\begin{eqnarray}
 f(T)=T+\xi \, T^2 \ ,\label{fT}
\end{eqnarray}
where $\xi$ is a real constant. Obviously, $\xi=0$ gives the Teleparallel equivalent of General Relativity.

In \cite{paperI} we discuss in detail the best set of equations that should be numerically solved to model stars, which include the equations for $P$ and $B'$, namely,
\begin{eqnarray}
P &=& \frac{{c^4}\, {\rm e}^{-B}}{8\pi {G} r^4}\Big\{{r}^{2} \left(1-{{\rm e}^{B}} \right) +2{{\rm e}^{B}}({{\rm e}^{-1/2\,B}}-1)^3(3{{\rm e}^{-1/2\,B}}+1)\xi+
\nonumber \\
&+& r\Big[r^2 +12({{\rm e}^{-1/2\,B}}-1)^2\xi\Big] A'+ 2r^2 \xi({{\rm e}^{-1/2\,B}}-1)(3{{\rm e}^{-1/2\,B}}-1) \, {A'}^2\Big\}
\label{press}
\end{eqnarray}

\par and

{
\begin{eqnarray} 
B'&=& -\frac{{{\rm e}^{B}}}{r}\bigg\{{r}^{4}\big(1-{{\rm e}^{-B}}-8\,\pi \,{r}^{2}\rho\big)
-96\pi\,{r}^{4}\rho ({{\rm e}^{-1/2\,B}}-1)^2\xi+ 
\nonumber \\    
&-& 6\,{r}^{2}({{\rm e}^{-1/2\,B}}-1)^3(5+3{{\rm e}^{-1/2B}})\xi
-8({{\rm e}^{-1/2\,B}}-1)^5(11+9{{\rm e}^{-1/2B}})\xi^2 +
\nonumber \\    
&-& 8\,{r}^{3}{{\rm e}^{-1/2\,B}}({{\rm e}^{-1/2\,B}}-1)\Bigl[8\,\pi \,{r}^{2}\rho
+({{\rm e}^{-1/2\,B}}-1)(2{{\rm e}^{-1/2\,B}}+1)\Bigr]\,A'\xi+
\nonumber \\    
&-& 16\, r\, {{\rm e}^{-1/2\,B}}({{\rm e}^{-1/2\,B}}-1)^4(9{{\rm e}^{-1/2\,B}}+5)\,A'\xi^2
+2\,r^4 {{\rm e}^{-B}}({{\rm e}^{-1/2\,B}}-1)^2  \,{A'}^2\xi+
\nonumber \\    
 &-& 8\,r^2 {{\rm e}^{-B}}({{\rm e}^{-1/2\,B}}-1)^3 (9{{\rm e}^{-1/2\,B}}-1)\,{A'}^2\, \xi^2\bigg\}
\bigg/ 
 \bigg\{ r^4+16\,{r}^{2}({{\rm e}^{-1/2\,B}}-1)^2\xi+
\nonumber \\   
&+& 48\,({{\rm e}^{-1/2\,B}}-1)^4\xi^2
+16\,{r}^{3}{{\rm e}^{-1/2\,B}}({{\rm e}^{-1/2\,B}}-1) A'\xi +
\nonumber \\   
&+& 96\,r\,{{\rm e}^{-1/2\,B}}({{\rm e}^{-1/2\,B}}-1)^3 A'\xi^2 
   +48\,{r}^{2}{{\rm e}^{-B}}({{\rm e}^{-1/2\,B}}-1)^2
 {A'}^2\xi^2 \bigg\} 
\label{BL}
\end{eqnarray}
}

From equations (\ref{026}) - (\ref{027}), one also obtains a differential equation for $A''$, but it is not necessary to use it in the numerical integration \cite{paperI}. Instead, the \emph{conservation equation} is used, namely,
\begin{equation}
    2 P' + (P+\rho)A' = 0  \ ,
\label{ce1}    
\end{equation} 
which, like in General Relativity, also holds in $f(T)$ gravity (see Ref. \cite{Bohmer} for details). 
Also, note that the energy density $\rho$ in the above equations is related to the pressure $P$ through a equation of state (EOS). 

Solving the system of equations (\ref{press}), (\ref{BL}) and (\ref{ce1}) for a given EOS and the appropriate boundary conditions, one obtains the radius of the star $R$, $P(r)$ and $\rho(r)$. Another obvious quantity that must be necessarily calculated is the mass $M$ of the star.  

Recall that in General Relativity the mass equation reads: 
\begin{eqnarray}
\frac{dM}{dr}=4\pi \rho  r^2\,
\label{dmdr}
\end{eqnarray}
which is just like the newtonian equation.
It is worth stressing that the mass of stars in General Relativity is a well-defined quantity, whereas, in alternative gravity, this is not necessarily the case \cite{olmo}. Therefore, there is space in the literature to keep addressing this issue.

The total rest mass $M_0$, although it is not an observable quantity, is an interesting quantity to be considered, since it can be used to compare different theories of gravity. For given EOS and central density, for instance, different theories provide different values of $M_0$. The total mass $M_0$ is obtained by the solution of the following differential equation
\begin{eqnarray}
\frac{dM_0}{dr}=4\pi \rho_0 \,e^{B/2} r^2
\label{dm0dr}
\end{eqnarray}
where $\rho_0$ is the rest mass density.

In the next section, we address such an important issue of calculating the mass of stars in $f(T)$ gravity and compare them with the one obtained via equation (\ref{dmdr}).

\section{Compact stars on $f(T)$ models and  the issue of mass}\label{models}

{Before considering the calculation of mass, it is worth recalling the basic prescription to model stars in $f(T)$ gravity. For further details we refer the reader to \cite{paperI}. As already mentioned, the calculation of mass does not affect the calculation of the radius of the star as well as its density  and pressure profiles, i.e., $\rho(r)$ and $P(r)$, respectively. Therefore, we start discussing how to solve equations (\ref{press}), (\ref{BL}) and (\ref{ce1}) for a given EOS.
}

{
The numerical integration is performed by means of a numerical code written in Python. To solve numerically equations (\ref{press}), (\ref{BL}) and (\ref{ce1}) for a given EOS, it is necessary to provide the following central boundary condition
\begin{equation}
    P = P_c \quad {\rm at} \quad r =0.
\end{equation}
}

{Note that, in comparison to General Relativity, we have an additional equation, namely, the differential equation for $B(r)$. So we need to provide a central boundary condition for B, namely, $B_c = 0$, whose choice has to do with regularity at $r = 0$.}

{
The radius $R$ of the star is the value of $r$ for which $P(r) = 0$. In practice, one starts the integration of the set of differential equations at $r=0$ and continues it till the value of $r$ for which $P(r)=0$.
}

{\subsection{The calculations of mass}}

{
Besides the use of equations (\ref{dmdr}) and (\ref{dm0dr}) to obtain $M$ (hereafter referred to as $M_{GR}$) and $M_0$, we now present two other ways to calculate mass in $f(T)$ gravity. Before proceeding, notice that the calculation of $M_0$ in $f(T)$ gravity is unambiguously given, since it only takes into account the total rest mass.  The internal energy and the gravitational potential energy do not enter in the calculation of $M_0$. On the other hand, in the calculation of $M_{GR}$, these forms of energy referred to above are taken into account. Since gravity is now modified, this is why there is no guarantee that $M_{GR}$ accounts for the mass of a compact object in $f(T)$ gravity.
}

{A way to calculate the mass has to do with the fact that the mass of the object should appear in some way in the metric it generates. That is, the mass should be encoded in $A(r)$ and $B(r)$ metric potentials. As a result, one can obtain the mass of an asymptotically Minkowskian metric ($M_{\infty}$), i.e., the ``mass measured by an observer at infinity''. As is well known, applying this prescription in General Relativity one obtains $M_{GR}$, since the mass that appears in the Schwarzschild metric is nothing but 
``mass measured by an observer at infinity''.
}

First, note that equation (\ref{BL}) also holds for vacuum, since $\rho$ smoothly goes to zero. Considering equation (\ref{BL}) for r $\rightarrow \infty$, one obtains
\begin{equation}
    B' = - \frac{B}{r} \qquad {\rm for \; r \rightarrow \infty},
\end{equation}
whose solution reads
\begin{equation}
    B(r) = \frac{C}{r}, \label{BA}
\end{equation}
where C is a positive constant. Far from the object, no terms involving $\xi$ appears. 

Equation (\ref{press}) also holds for vacuum, since $P$ smoothly goes to zero.  Now, considering equation (\ref{press}) for $r$ $\rightarrow \infty$ and using (\ref{BA}), one obtains
\begin{equation}
    A' = \frac{B}{r} = \frac{C}{r^2} \qquad {\rm for \; r \rightarrow \infty},
\end{equation}
whose solution reads
\begin{equation}
    A(r) = - \frac{C}{r}, \label{AA}
\end{equation}
where we consider that $A(r \rightarrow \infty ) = 0$ to set the other integration constant equal to zero. Again, no terms in $\xi$ appear in the asymptotic solution for $A(r)$.

The metric given by equation (\ref{metric})
for r $\rightarrow \infty$ reads
\begin{equation}
    ds^2=(1+A(r))\, dt^2-(1+B(r))\, dr^2-r^2\, d\theta^2-r^2 \sin ^2 \theta \, d\phi^2 \label{metric1},
\end{equation}
which can be rewritten, with the use of equations (\ref{BA}) and (\ref{AA}), as follows
\begin{equation}
    ds^2=\left(1- \frac{C}{r}\right)\, dt^2-\left(1 + \frac{C}{r}\right)\, dr^2-r^2\, d\theta^2-r^2 \sin ^2 \theta \, d\phi^2 \label{metric2}.
\end{equation}

Equation (\ref{metric2}) strongly suggests that $C$ is related to the ``mass measured by an observer at infinity'' ($M_\infty$). Consequently, we set $C \equiv 2 M_\infty$. This is just like Schwarzschild metric very far from the source.

In practice, to calculate $M_\infty$ it is necessary to solve equations (\ref{press}) and (\ref{BL}) for vacuum, starting from the star's surface till, say, hundred or thousand times the star's radius. This ensures that equation (\ref{BA}) holds, which is corroborated by the numerical results. 
Therefore $M_\infty = C/2$ is easily obtained, namely,
\begin{equation}
    M_{\infty} = \frac{C}{2} =  \lim_{r\to\infty} \frac{1}{2\, }r\,B(r).
\end{equation}

As already mentioned, the radius of the star does not depend on the calculation of mass, no matter what definition is used for its calculation.
 
It is worth noticing that an observer near to the star feels the contribution of $M_\infty$ and additional terms that come from the $T^2$ term.

Another way to calculate the mass, which we call $M_S$, is obtained by considering that the metric for vacuum is given by the Schwarzschild metric.
In Refs. \cite{Ganiou,Kpa}, for example, this procedure is adopted. Although it is important to bear in mind that such a procedure is misleading, because the external spacetime of a spherical star in $f(T)$ gravity is not given by the 
Schwarzschild metric \cite{Ilijic3}. In any case, we will calculate the mass using this procedure in order to compare it with the other ways of calculating the mass of the star in this article.

To proceed, we then write
\begin{equation}
    e^{-B(r)} = 1 - 2\,m_S(r)/r \qquad {\rm at } \; r = R.
\end{equation}
Consequently the ``Schwarzschild mass'' then reads
\begin{equation}
    M_S = m_S(R) = \frac{R}{2} (1 - e^{-B(R)}).
\end{equation}

In what follows we present and compare a series of calculations for the masses $M_{GR}$, $M_\infty$ and $M_S$ for FPS and SLY equations of states (EOSs)\footnote{Tables with FPS and SLY EOSs can be found, e.g., at \url{http://xtreme.as.arizona.edu/NeutronStars/}} {for neutron stars}. It is worth noting that $M_0$ is also included in some of the figures presented below.

\begin{figure}
    \centering
    \includegraphics[width=0.48\linewidth]{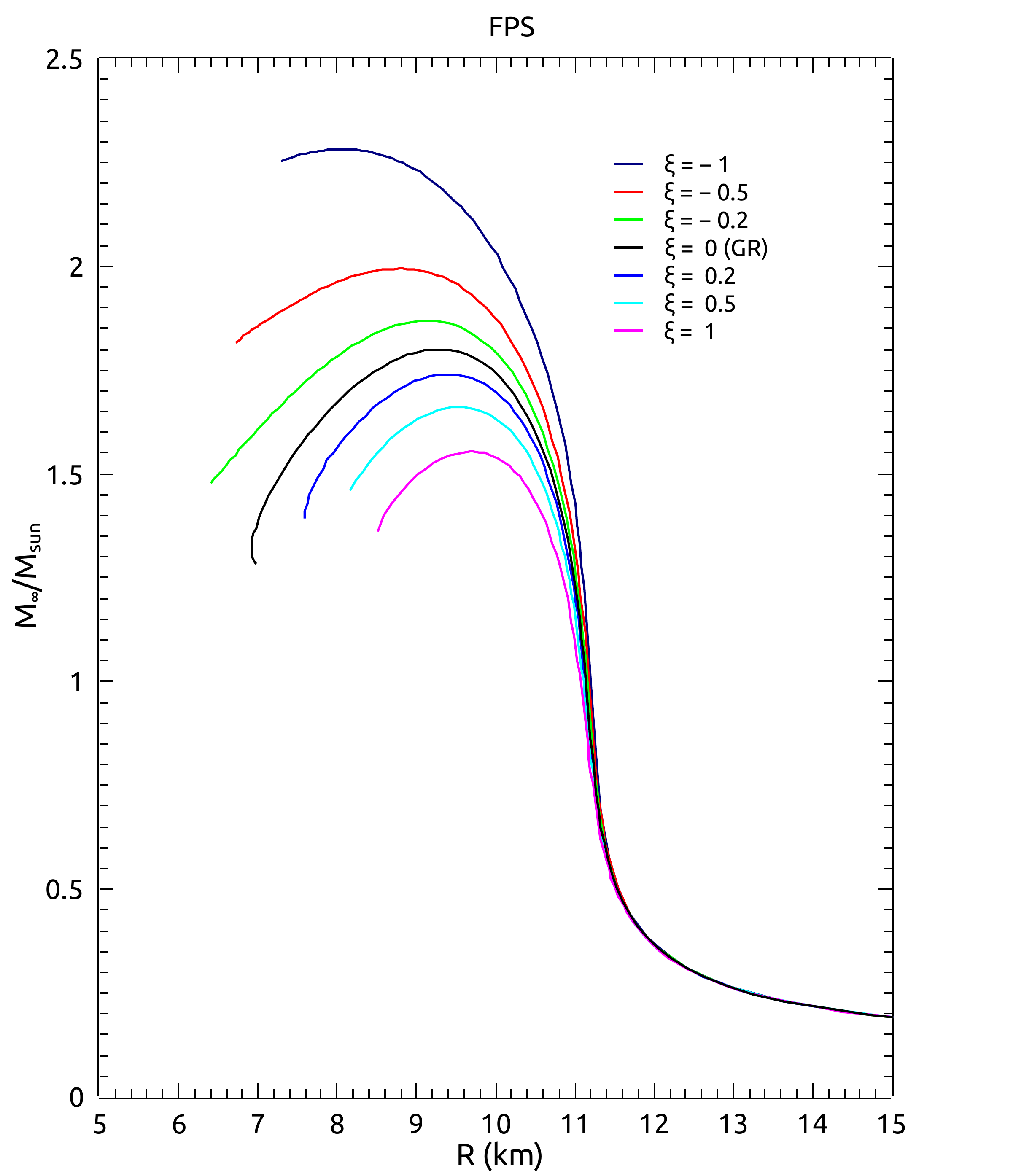}
    \centering
    \includegraphics[width=0.49\linewidth]{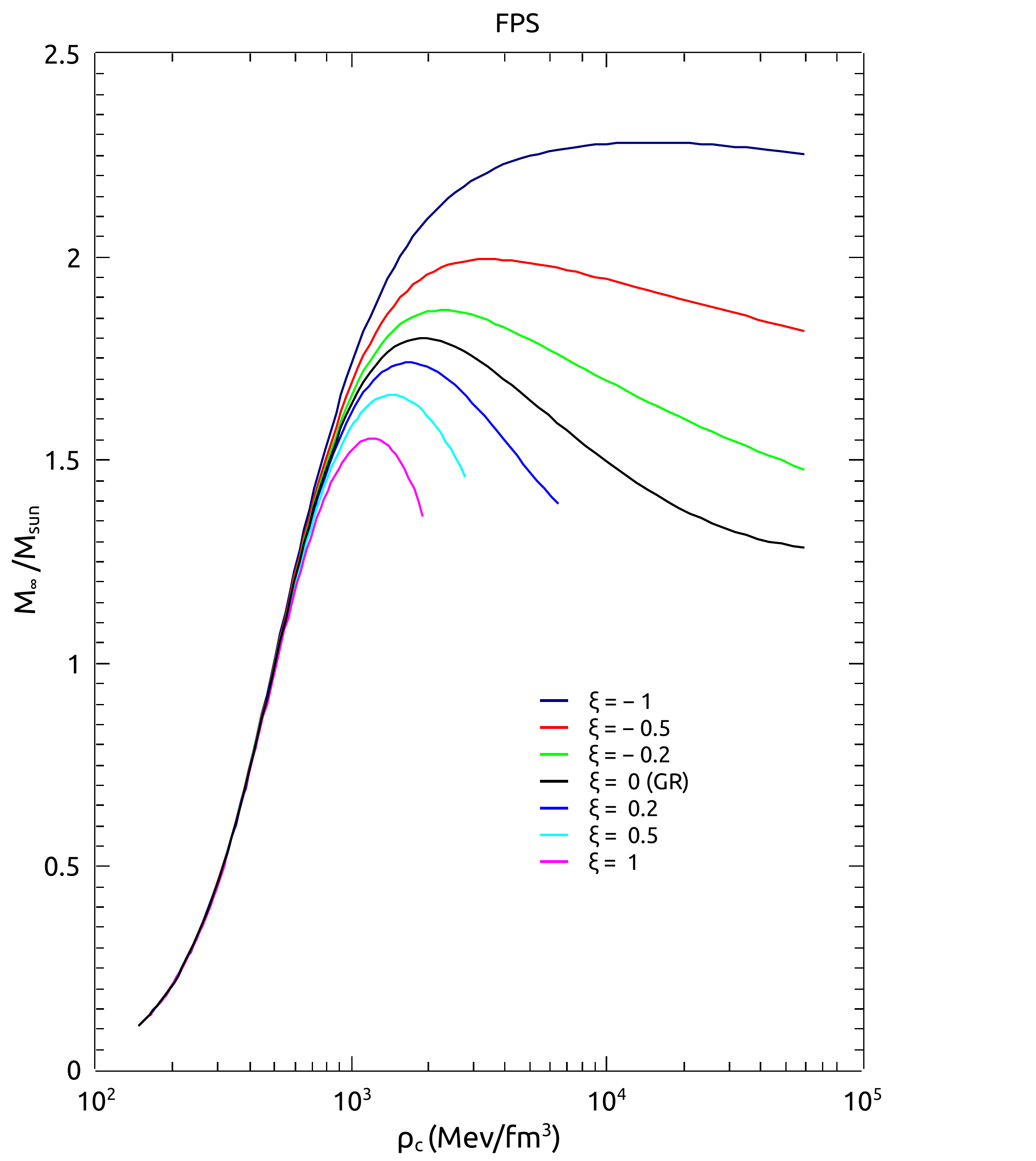}
    \caption{Sequences of mass M$_\infty$ as a function of the radius R (left panel) and the central energy density $\rho_c$ (right panel) for FPS EOS and different values of $\xi$, which is given in units of the square of the gravitational radius of the Sun $r_{gs}^2= 4G^2M_\odot^2/c^4$.}
    \label{FPSMRD}
\end{figure}

Since $M_\infty$ is in fact the most appropriate way to calculate mass in $f(T)$ gravity, we start presenting ``$M_\infty$ $\times$ Radius'' and ``$M_\infty$ $\times$ $\rho_c$'' sequences for a given EOS. In figure \ref{FPSMRD} we present the referred sequences for FPS EOS and different values of $\xi$, which is given in units of the square of the gravitational radius of the Sun ($r_{gs} = 2GM_\odot/c^2 \simeq 2.95\, {\rm km})$. 

As a general conclusion, the different values of $\xi$ considered  here show that the sequences differ significantly from that of General Relativity. Note that there is a maximum mass for each value of $\xi$ considered, and the lower $\xi$, the greater the maximum mass. Furthermore, the lower $\xi$, the grater the compactness. 

We now compare $M_{GR}$, $M_S$ and $M_\infty$ for the FPS EOS and $\xi =  \pm 1$ $r_{gs}^2$. It is worth noting that $M_0$ is included only as a reference value, since it can not represent the mass of a compact object. Recall that the mass of a bound configuration must be lower than $M_0$.

\begin{figure}[!ht]
        \centering
\includegraphics[width=0.48\linewidth]{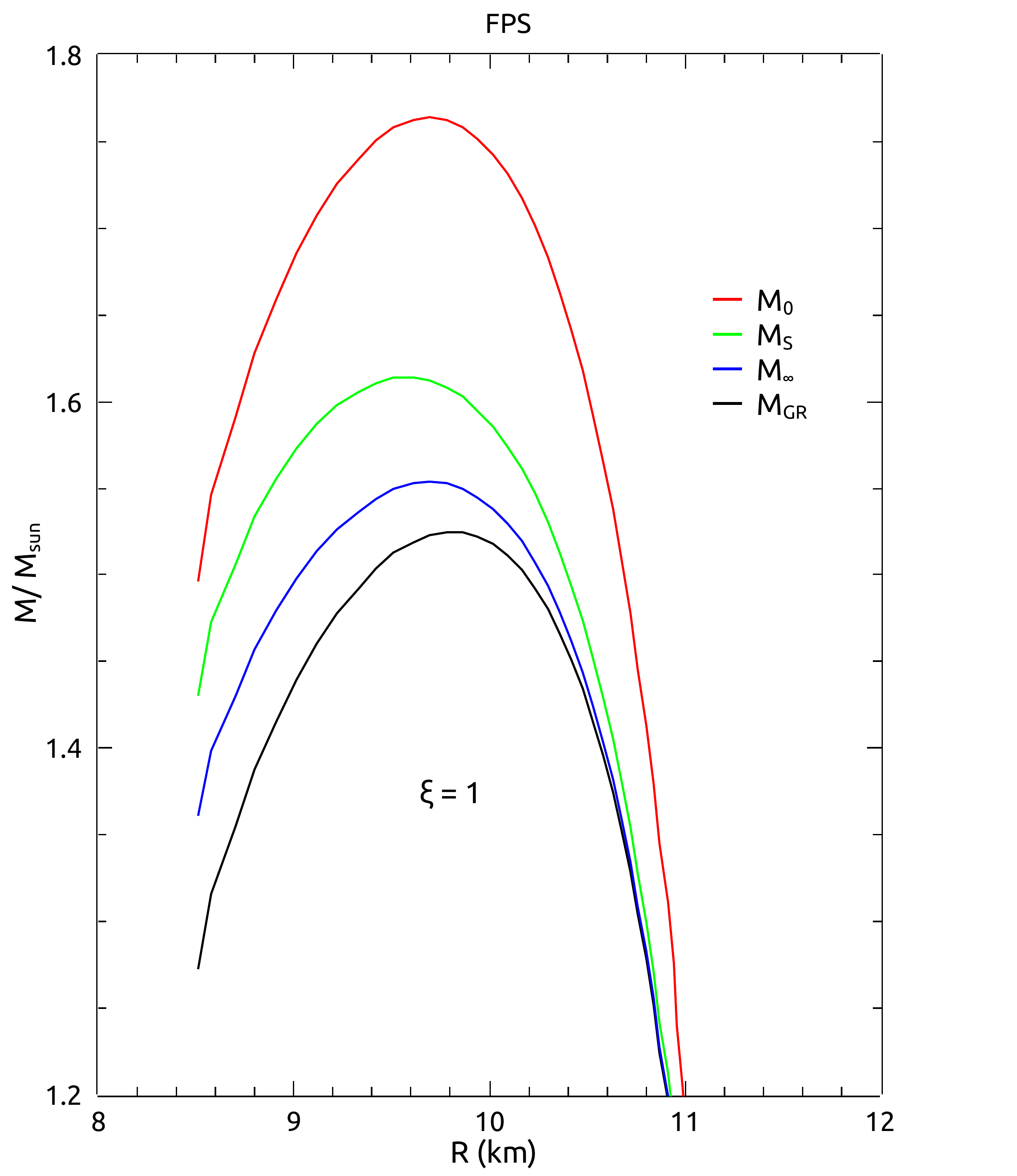}
        \centering
\includegraphics[width=0.48\linewidth]{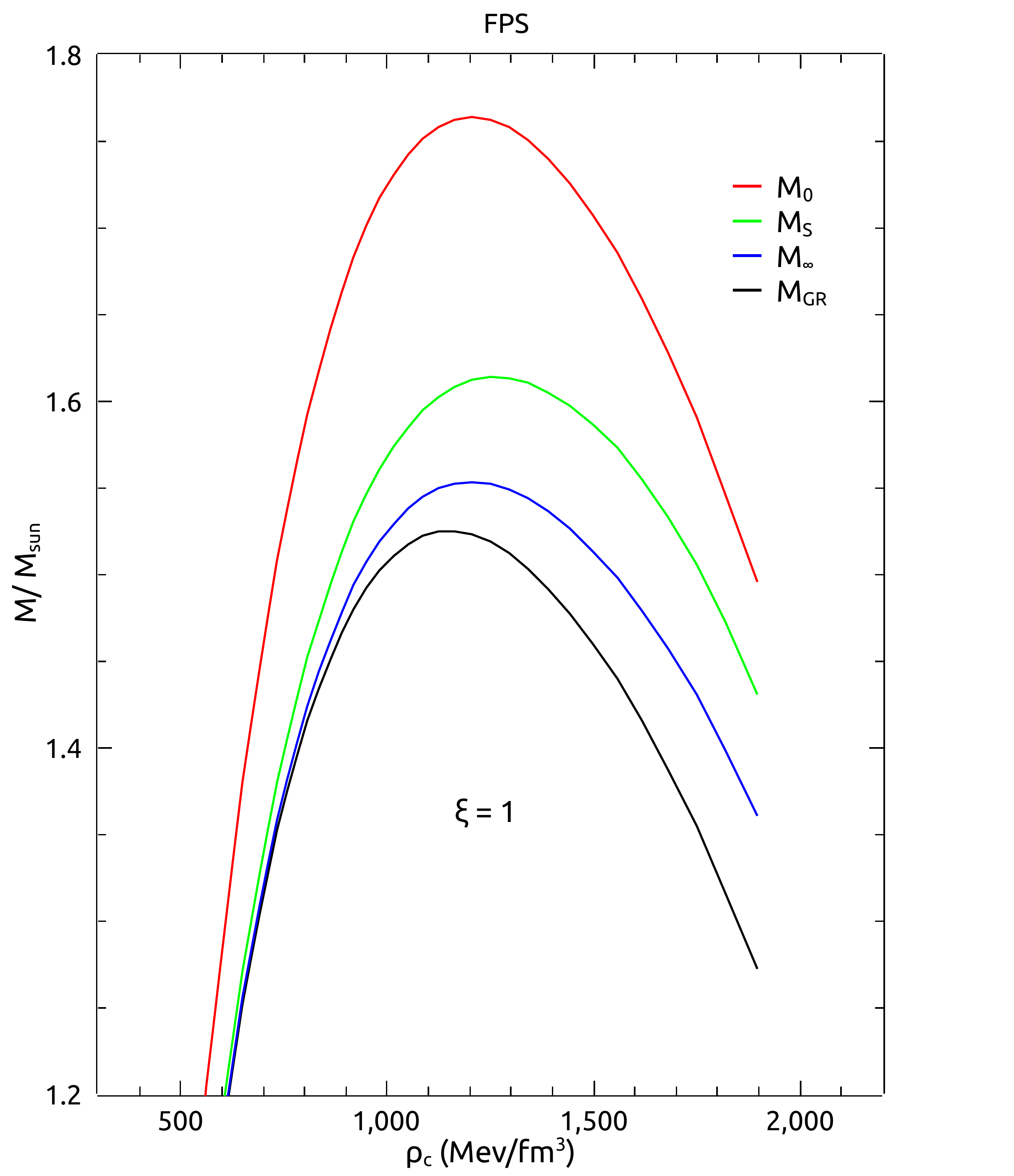}
        \caption{Sequences of $M_{GR}$, $M_S$, $M_\infty$ and $M_0$ as a function of the radius R (left panel) and the central energy density $\rho_c$ (right panel) for FPS EOS and  $\xi = 1$  $r_{gs}^2$.}
    \label{FPS10P}
\end{figure}

Figure \ref{FPS10P} shows sequences for $M_{GR}$, $M_S$, $M_\infty$ and $M_0$ as a function of the radius R (left panel) and the central energy density $\rho_c$ (right panel) for FPS EOS and  $\xi = 1$  $r_{gs}^2$. First, note that the curves for $M_{GR}$, $M_S$ and $M_\infty$ are below the curve for $M_0$. This is an indication that the stars are bound, no matter what definition of mass is adopted.
Second, the curves for $M_{GR}$, $M_S$ and $M_\infty$ are almost the same till $\simeq$ 1.4 M$_\odot$. The maximum masses and their corresponding radii are clearly different.

Although not shown, our calculations indicate that the larger $\xi$ the closer the curves for $M_{GR}$, $M_S$ and $M_\infty$ become. 

\begin{figure}[!ht]
        \centering
        \includegraphics[width=0.48\linewidth]{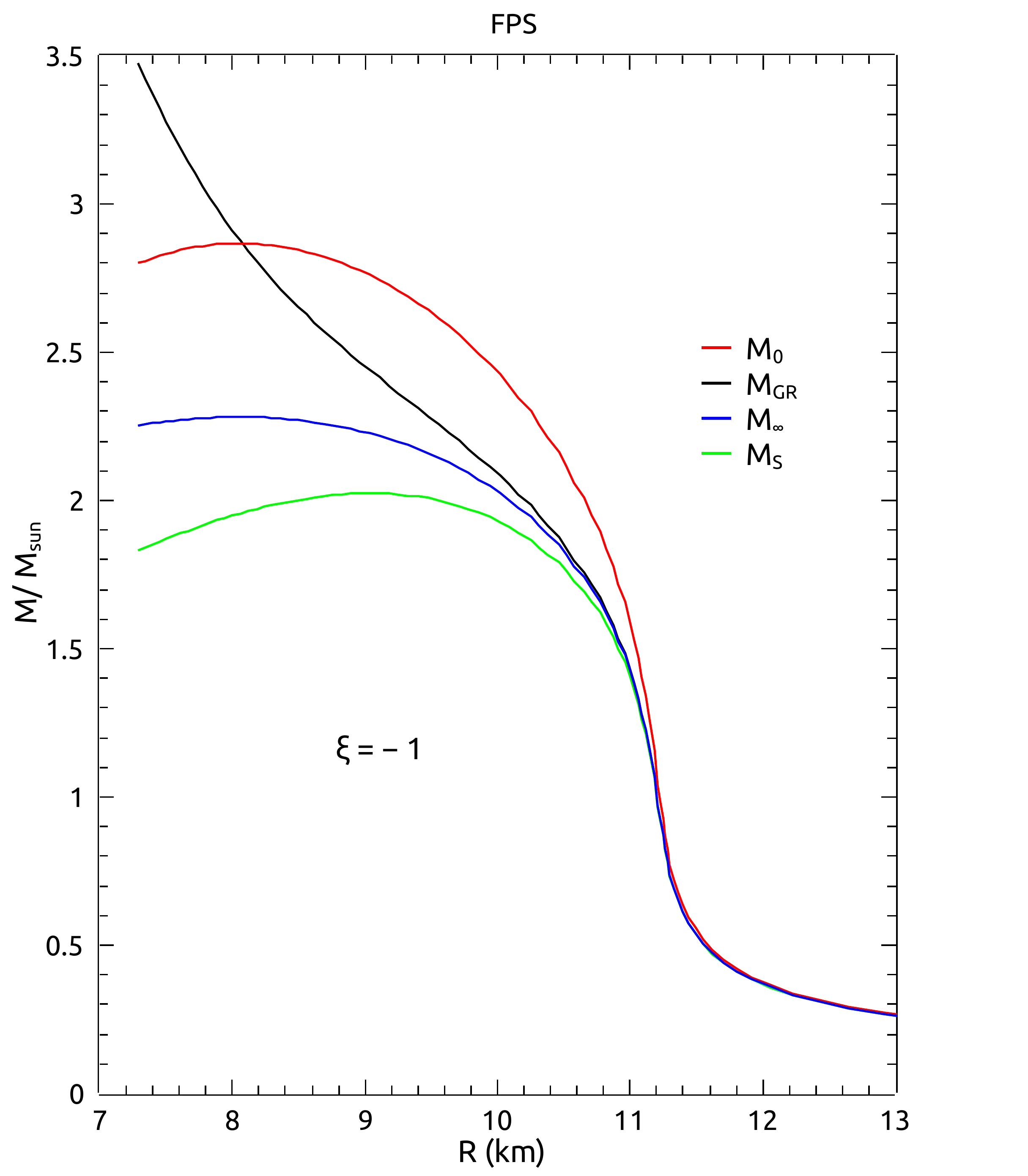}
        \centering
        \includegraphics[width=0.48\linewidth]{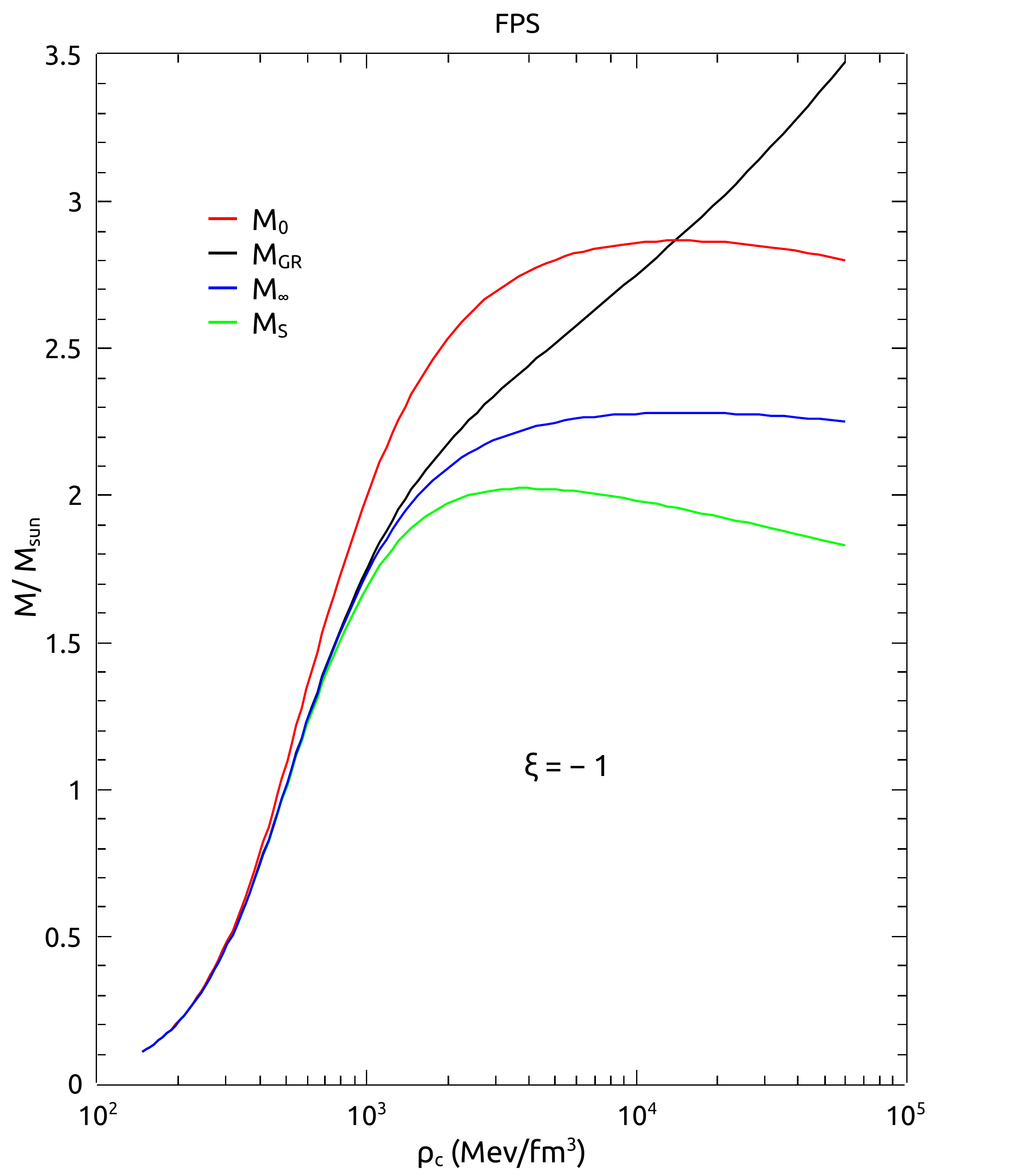}
        \caption{Sequences of $M_{GR}$, $M_S$, $M_\infty$ and $M_0$ as a function of the radius R (left panel) and the central energy density $\rho_c$ (right panel) for FPS EOS and  $\xi = - 1$  $r_{gs}^2$.}
    \label{FPS10M}
\end{figure}

Let us now compare the different definitions of masses for $\xi = - 1$ $r_{gs}^2$. As can be seen in figure \ref{FPS10M}, the curves for $M_{GR}$, $M_S$ and $M_\infty$ are almost the same till $\simeq$ 1.4 M$_\odot$. For higher masses the curves are significantly different. Note also that $M_{GR}$ > $M_0$ for energy densities higher than $\sim 10^4$ Mev/fm$^3$, which would indicate that the configuration is unbound. This could well be an indication that $M_{GR}$ would be a ill-defined mass of a star in $f(T)$ gravity.

Recall that in GR the maximum compactness of a spherical star (Buchdahl limit) is M/R = 4/9 (in geometrical units). In alternative gravity this is not necessarily the case, but it serves as a reference value. Adopting $M_{GR}$, one sees that negative values of $\xi$ can yield compactness that far exceeds 4/9. This could also be an indication that $M_{GR}$ does not give appropriately the mass of compact objects in $f(T) = T + \xi T^2$.

\begin{figure}
    \centering
    \includegraphics[width=0.48\linewidth]{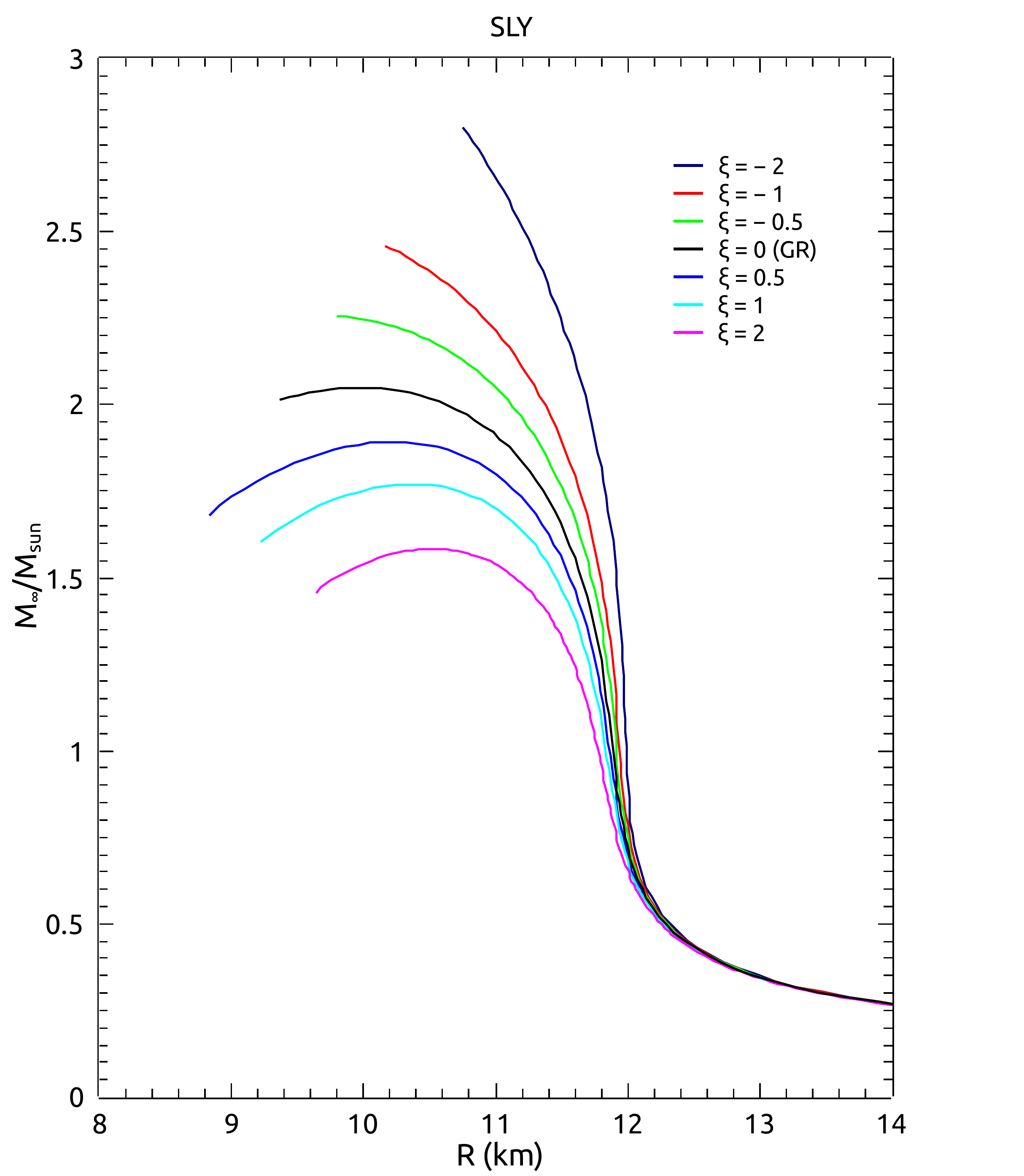}
    \centering
    \includegraphics[width=0.49\linewidth]{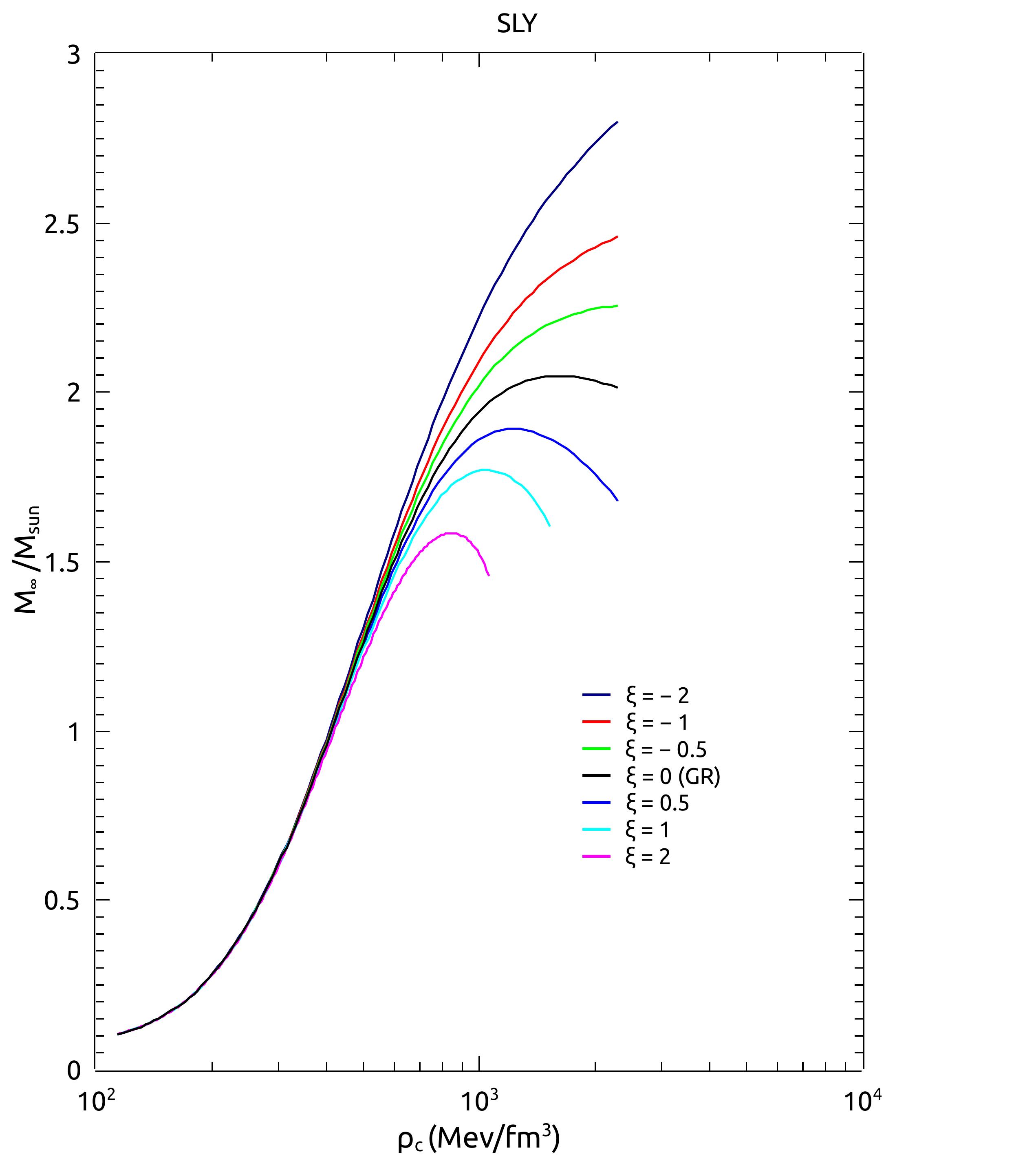}
    \caption{Sequences of mass M$_\infty$ as a function of the radius R (left panel) and the central energy density $\rho_c$ (right panel) for SLY EOS and different values of $\xi$, which is given in units of the square of the gravitational radius of the Sun $r_{gs}^2= 4G^2M_\odot^2/c^4$.}
    \label{SLYMRD}
\end{figure}

\begin{figure}[!ht]
        \centering
        \includegraphics[width=0.48\linewidth]{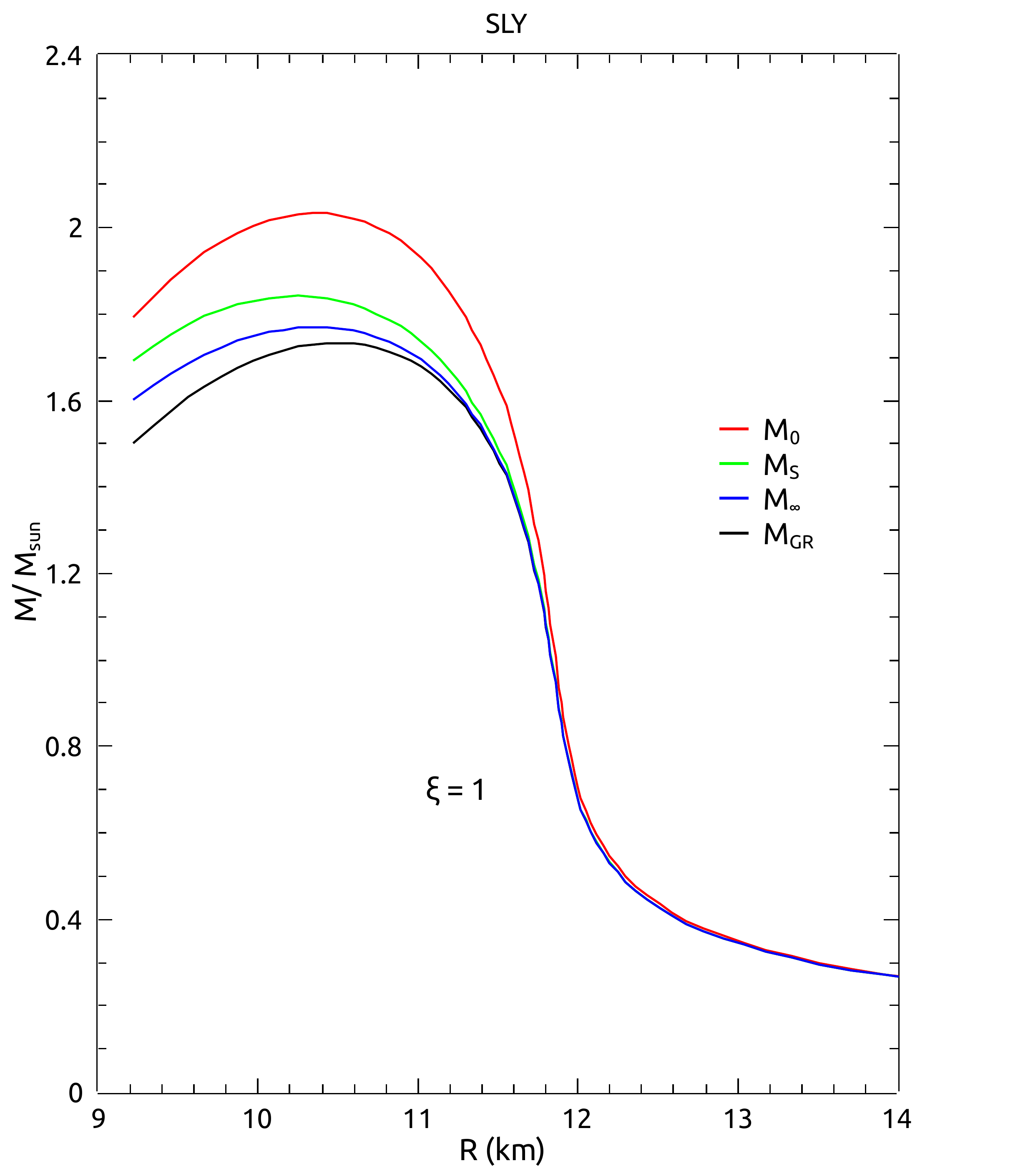}
        \centering
        \includegraphics[width=0.49\linewidth]{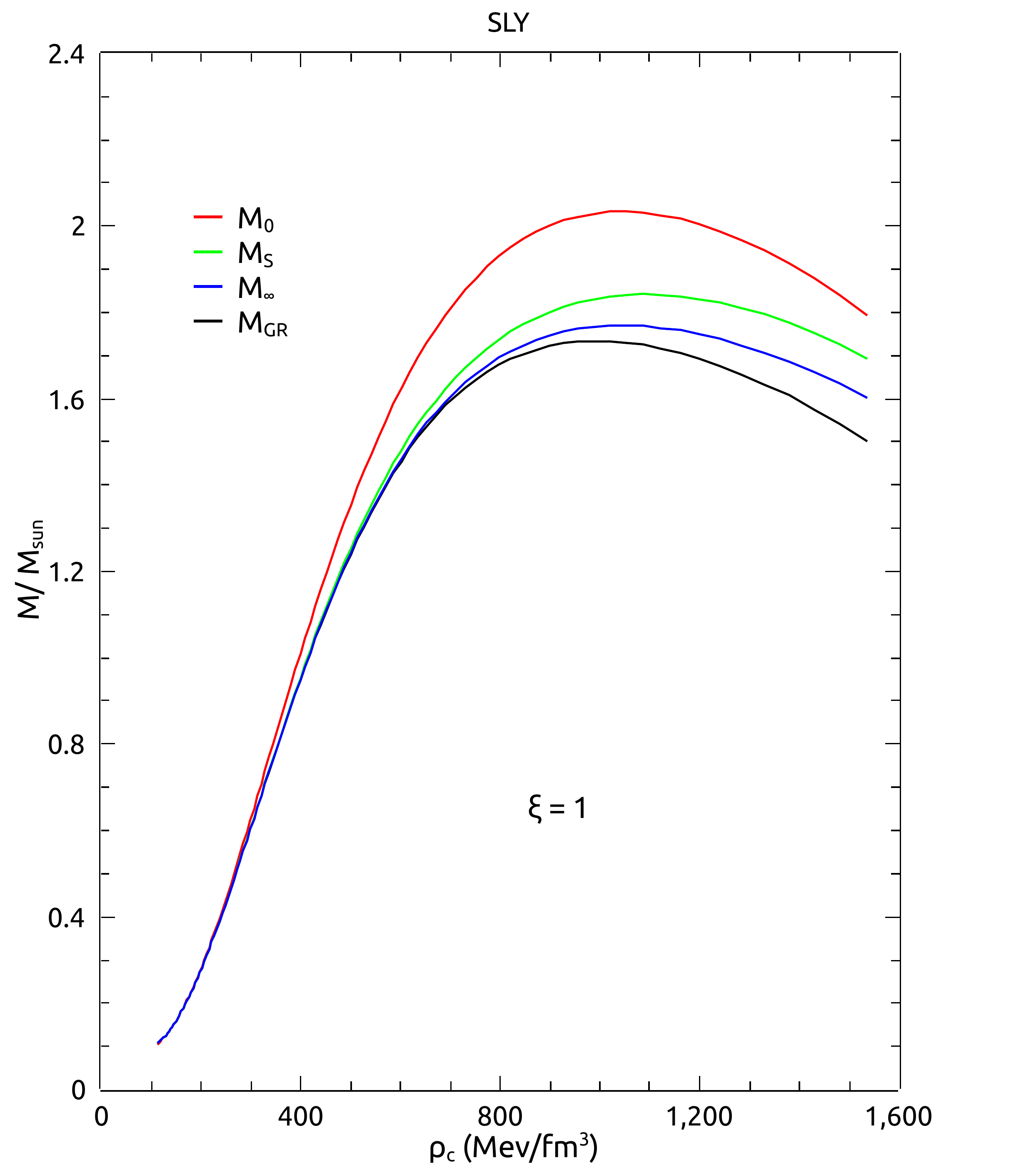}
        \caption{Sequences of $M_{GR}$, $M_S$, $M_\infty$ and $M_0$ as a function of the radius R (left panel) and the central energy density $\rho_c$ (right panel) for SLY EOS and  $\xi = 1$  $r_{gs}^2$.}
    \label{SLY10P}
\end{figure}

\begin{figure}[!ht]
        \centering
        \includegraphics[width=0.47\linewidth]{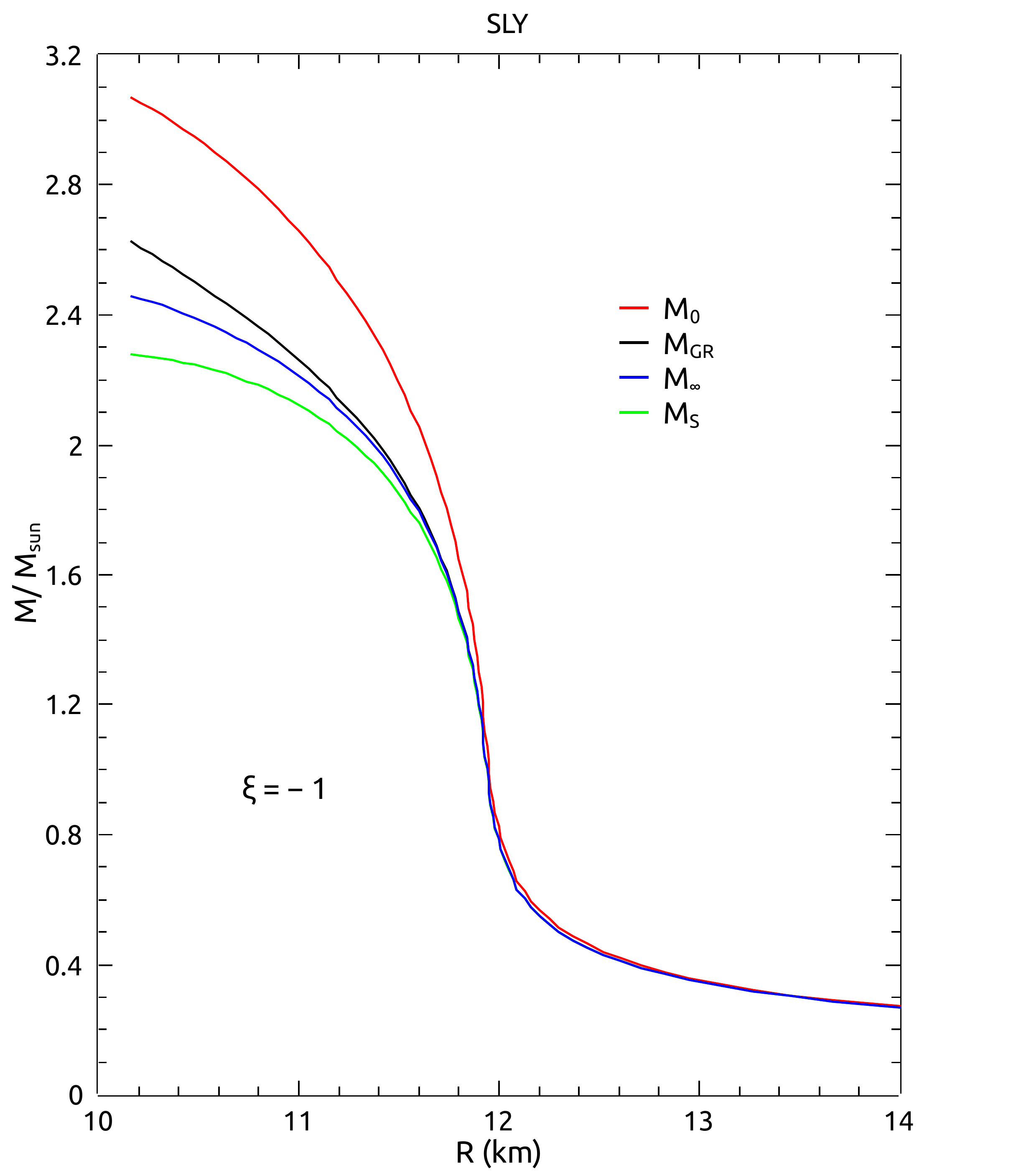}
        \centering
        \includegraphics[width=0.49\linewidth]{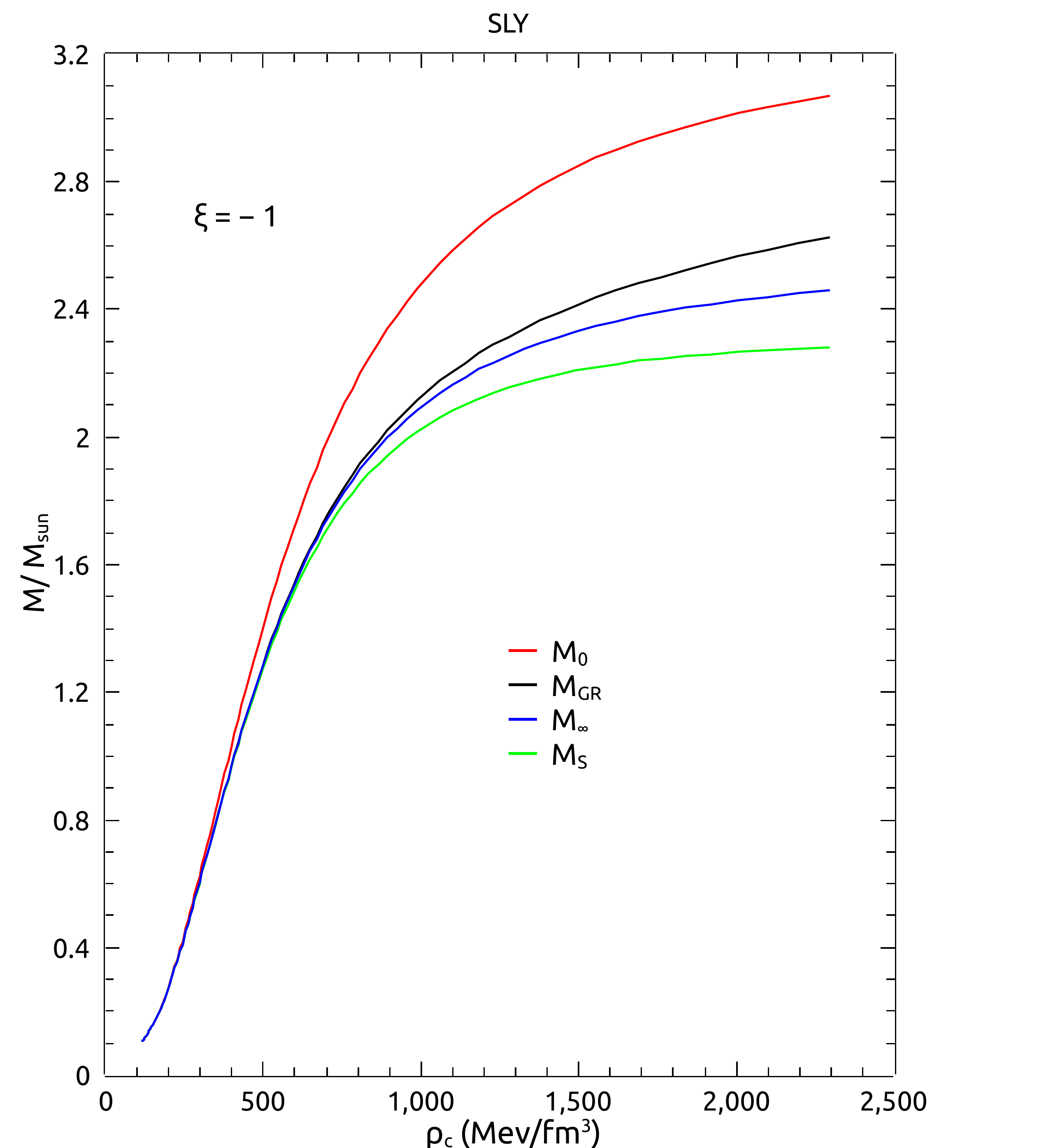}
        \caption{Sequences of $M_{GR}$, $M_S$, $M_\infty$ and $M_0$ as a function of the radius R (left panel) and the central energy density $\rho_c$ (right panel) for SLY EOS and  $\xi = - 1$  $r_{gs}^2$.}
    \label{SLY10M}
\end{figure}

To see how an EOS other than FPS behaves under different mass definitions, we also present calculations for the SLY EOS. Figures \ref{SLYMRD}, \ref{SLY10P} and \ref{SLY10M} show the results of our calculations.


{Besides the differences due the mass definition explored in this paper, one interesting fact regarding to the the torsion scalar was addressed in \cite{paperI}. While in GR the Ricci scalar outside (vacuum) of any matter distribution is zero, the torsion $T$ is not null outside a spherically symmetrical distribution. On the other hand, at the center of a star, the Ricci scalar is not zero and, instead, the torsion scalar is. Additionally, $T(r)$ goes to zero outside the star at a distance of only a few radii from the star. For example, for $\xi=-0.1$ and $\rho_c$ = 1, we obtain  $T(3R) \sim T(R)/100$. Thus, the spacetime becomes almost flat relatively near the star surface as in GR.}

{\subsection{Other approaches for calculations of mass}}


{As already mentioned, the definition of the mass of the star in $f(T)$ gravity is considered by many authors as an open problem, see \cite{olmo,Bahamonde} for a brief discussion. In GR, the calculation of the mass of the compact spherically symmetric object can be obtained directly from
the merging of the interior with the exterior Schwarzschild metric and this definition coincides with the ADM mass of the spacetime \cite{ADM}. This is not so for $f(T)$ gravity, since there is not the vacuum metric in a closed form. Therefore, it is natural to find different approaches in the literature regarding this subject, such as, considering the same mass definition of GR despite of the ambiguity, obtaining the mass from the asymptotic behavior of the metric at $r\rightarrow \infty$, calculating the total particle number, etc.}

{In order to exemplify this ambiguity, let us cite some recent works. Some authors basically reproduce the calculation from GR given in (\ref{dmdr}). For example, \cite{Nashed} presents charged anisotropic solution for pulsars in the framework of TEGR using a generalisation of (\ref{dmdr}); others use exactly the same expression for $\frac{dm}{dr}$, such as \cite{Pace,Zubair,Lin}. In \cite{Ilijic}, the authors compute the total particle number $N$ by integrating the particle number density $n = dN/dV$ over the interior of the star. In \cite{Ilijic3}, they solve perturbatively the spherically symmetric vacuum gravitational equations around the Schwarzschild solution.}

{In summary, the purpose here was to bring into focus a subject sometimes neglected in recent papers and comparing qualitatively the different approaches, since different equations of state are used and, therefore, a direct comparison is not suitable.}
\\

\section{{Final Remarks}}\label{conclusions}

Modeling stars has been an important way to probe alternative theories of gravity. However, unlike in General Relativity where the mass of an spherical object is a well-defined quantity, in alternative theories this does not necessarily happen.

In this article we consider this important issue of calculating mass in the particular $f(T) = T + \xi\, T^2$ gravity. In this alternative gravity the issue related to the definition of mass is considered by some authors to be still an open problem.

By considering four different masses ($M_0$, $M_{GR}$, $M_\infty$ and $M_S$), we have seen that only two of them have in fact physical sense, namely, $M_0$ and $M_\infty$. Notice that, in General Relativity, $M_{GR}$, $M_\infty$ and $M_S$ are identical. On the other hand, for the model considered in the present paper, these masses can be very different, depending on the value of $\xi$.

For $M_0$, as we have already mentioned previously, it gives the amount of mass in particles of a star, which does not include neither the internal energy nor the gravitational energy. Although it is not an observable quantity, it can well be used to compare different gravity theories. 

Concerning $M_{GR}$, the problem possibly has to do with the fact that the effect of the term $\xi \, T^2$ is not taken into account properly.

For $M_S$, the main problem has to do with the assumption that the spacetime exterior to the spherical star is given by the Schwarzschild metric. This is an incorrect assumption, since the spacetime exterior to a spherical star is not known in closed form for $f(T)$ theories and it is certainly not given by the Schwarzschild metric.

{Finally, we reaffirm that the most suitable way to calculate mass is to consider $M_\infty$ due to its very definition, i.e., the mass measured by an observer at infinity. 
Therefore, $M_\infty$ seems to represent properly the mass of a spherical star in $f(T)$ gravity.}

\begin{acknowledgements}
J.C.N.A. thanks FAPESP $(2013/26258-4)$ and CNPq (308367/2019-7) for partial financial support. The authors would like to thank Rafael da Costa Nunes for discussions related to the $f(T)$ theory. {Last but not least, we thank the referee for the valuable review of our article, which helped to improve it.}

\end{acknowledgements}

\bibliographystyle{spphys}       

\begin{thebibliography}{99}
\bibitem{olmo} G. J. Olmo, D. Rubiera-Garcia, A. Wojnar, \emph{Stellar structure models in modified theories of gravity: Lessons and challenges}, \emph{Phys. Report} {\bf 876} (2020) 1.

\bibitem{TT}Y. Cai et al., \emph{$f(T)$ teleparallel gravity and cosmology}, \emph{Rept. Prog. Phys.} {\bf 79} (2016), 106901.

\bibitem{Ferraro} R. Ferraro, F. Fiorini, \emph{Modified teleparallel gravity: Inflation without inflaton}, \emph{Phys. Rev. D} {\bf 75}, (2007) 084031 [gr-qc/0610067].

\bibitem{Bahamonde} S. Bahamonde et al. \emph{Teleparallel Gravity: From Theory to Cosmology. Rep. Prog. Phys. 86} (2023) 026901 (207pp)

\bibitem{Bohmer} 
C. G. Böhmer, A. Mussa and N. Tamanini, \emph{Existence of relativistic stars in f(T) gravity}, \emph{Class. Quantum Grav.} {\bf 28} (2011) 245020.

\bibitem{Ilijic} S. Iliji\'c, M. Sossich, \emph{Compact stars in $f(T)$ extended theory of gravity}, \emph{Phys. Rev. D} {\bf 98} (2018) 064047.

\bibitem{Ilijic20} S. Iliji\'c, M. Sossich, \emph{Boson stars in $f(T)$ extended theory of gravity}, \emph{Phys. Rev. D} {\bf 102} (2020) 084019.

\bibitem{paperII} 	J.C.N. de Araujo, H.G.M. Fortes, Solving Tolman-Oppenheimer-Volkoff Equations in f(T) Gravity: a Novel Approach Applied to Polytropic Equations of State. Braz J Phys 53, 75 (2023). https://doi.org/10.1007/s13538-023-01293-x.

\bibitem{paperIII} J. C. N. de Araujo, H. G. M. Fortes, Solving Tolman-Oppenheimer-Volkoff equations in f(T) gravity: a novel approach applied to some realistic equations of state. International Journal of Modern Physics D, Vol. 31, No. 13, 2250101 (2022).

\bibitem{Vilhena} S.G. Vilhena et al, Neutron stars 434 on modified teleparallel gravity. JCAP04(2023)044. https://doi.org/10.1088/1475-7516/2023/04/044.

\bibitem{Ganiou} M. G. Ganiou et al., \emph{Strong magnetic field effects on neutron stars within f(T) theory of gravity}, \emph{Eur. Phys. J. Plus} {\bf 132} (2017) 250.


\bibitem{Nashed} G. G. L. Nashed, S. Capozziello, Stable and self-consistent compact star models in teleparallel gravity Eur. Phys. J. C (2020) 80:969

\bibitem{Sharif} M. Sharif and A. Waseem, Physical behavior of anisotropic compact stars in $f(R, T, R_{\mu\nu}T^{\mu\nu})$ gravity, Astrophys Space Sci (2016) 361:27.

\bibitem{Zubair}M. Zubair, A. Ditta, G. Abbas and R. Saleem, Physical aspects of
anisotropic compact stars in $f(T,\mathcal{T})$ gravity with off diagonal tetrad, Chinese Physics C Vol. 45, No. 8 (2021) 085102. 

\bibitem{Saleem} R. Saleem, F. Kramat, M. Zubair, Interior solutions of compact stars in gravity under Karmarkar condition, Physics of the Dark Universe 30 (2020) 100592.

\bibitem{Pace} M. Pace and J. L. Said, Quark stars in $f(T,\mathcal{T})$-gravity, Eur. Phys. J. C (2017) 77:62.

\bibitem{Ilijic3}A. DeBenedictis and S. Ilijic, Spherically symmetric vacuum in covariant $F(T) = T + \frac{\alpha}{2}
T^2 + O(T^\gamma)$ gravity theory, Phys. Rev. D 94 (2016)
124025 [arXiv:1609.07465 [gr-qc]].





\bibitem{ADM} R. Arnowitt, S. Deser, C. W. Misner, Republication of: The dynamics of general relativity. Gen Relativ Gravit 40, 1997–2027 (2008). https://doi.org/10.1007/s10714-008-0661-1.https://doi.org/10.1007/s10714-008-0661-1.


\bibitem{TOV} C. M. Will, \emph{Theory and experiment in gravitational physics}, Cambridge University Press (1993).

\bibitem{Atazadeh}K. Atazadeh and Misha Mousavi, \emph{Vacuum spherically symmetric solutions in f(T) gravity}, Eur. Phys. J. C73 2272 (2013)

\bibitem{paperI} H. G. M. Fortes and  J. C. N. de Araujo, \emph{Solving Tolman-Oppenheimer-Volkoff equations in f(T) gravity: a novel approach}, Class. Quantum Grav. {\bf 39} (2022) 245017, arXiv:2105.04473 [gr-qc].





\bibitem{Kpa}A. V. Kpadonou, M. J. S. Houndjo, M. E. Rodrigues, \emph{Tolman-Oppenheimer-Volkoff Equations and their implications for the structures of relativistic Stars in f(T) gravity}, \emph{Astrophys. Space Sci.} {\bf 361} (2016) 244.

\bibitem{Lin} R. Lin, X. Chen, X. Zhai, Realistic neutron star models in $f(T)$ gravity, Eur. Phys.
J. C (2022) 82:308


































\end{thebibliography}

\end{document}